\documentclass[aps,pra,twocolumn,superscriptaddress]{revtex4-1}
\usepackage{amsfonts}
\usepackage{amssymb}
\usepackage{amsmath}
\usepackage{epsfig}
\usepackage{color}
\usepackage{graphics, graphicx}
\usepackage{bbold}
\usepackage{psfrag}
\usepackage{mathcomp}
\usepackage{subfigure}
\usepackage{verbatim}
\usepackage[colorlinks,citecolor=blue]{hyperref}

\makeatletter

\newcommand{\Rmnum}[1]{\expandafter\@slowromancap\romannumeral #1@}
\makeatother

\begin{document}

\title{Robust quantum state transfer via topological edge states in superconducting qubit chains}
\author{Feng Mei}
\affiliation{State Key Laboratory of Quantum Optics and Quantum Optics Devices, Institute
of Laser Spectroscopy, Shanxi University, Taiyuan, Shanxi 030006, China}
\affiliation{Collaborative Innovation Center of Extreme Optics, Shanxi
University,Taiyuan, Shanxi 030006, China}
\author{Gang Chen}
\email{chengang971@163.com}
\affiliation{State Key Laboratory of Quantum Optics and Quantum Optics Devices, Institute
of Laser Spectroscopy, Shanxi University, Taiyuan, Shanxi 030006, China}
\affiliation{Collaborative Innovation Center of Extreme Optics, Shanxi
University,Taiyuan, Shanxi 030006, China}
\author{Lin Tian}
\email{ltian@ucmerced.edu}
\affiliation{School of Natural Sciences, University of California, Merced, California
95343, USA}
\author{Shi-Liang Zhu}
\email{slzhu@nju.edu.cn}
\affiliation{National Laboratory of Solid State Microstructures, School of Physics,
Nanjing University, Nanjing 210093, China}
\affiliation{Guangdong Provincial Key Laboratory of Quantum Engineering and Quantum
Materials, SPTE, South China Normal University, Guangzhou 510006, China}
\affiliation{Synergetic Innovation Center of Quantum Information and Quantum Physics,
University of Science and Technology of China, Hefei 230026, China}
\author{Suotang Jia}
\affiliation{State Key Laboratory of Quantum Optics and Quantum Optics Devices, Institute
of Laser Spectroscopy, Shanxi University, Taiyuan, Shanxi 030006, China}
\affiliation{Collaborative Innovation Center of Extreme Optics, Shanxi
University,Taiyuan, Shanxi 030006, China}
\date{\today }

\begin{abstract}
Robust quantum state transfer (QST) is an indispensable ingredient in scalable quantum information processing. Here we present an experimentally feasible mechanism for realizing robust QST via topologically protected edge states in superconducting qubit chains. Using superconducting Xmon qubits with tunable couplings, we construct generalized Su-Schrieffer-Heeger models and analytically derive the wave functions of topological edge states. We find that such edge states can be employed as a quantum channel to realize robust QST between remote qubits. With a numerical simulation, we show that both single-qubit states and two-qubit entangled states can be robustly transferred in the presence of sizable imperfections in the qubit couplings. The transfer fidelity demonstrates a wide plateau at the value of unity in the imperfection magnitude. This approach is general and can be implemented in a variety of quantum computing platforms.
\end{abstract}

\maketitle

\section{Introduction}

To realize large-scale quantum information processing, quantum states need
to be coherently transferred between distant nodes in a quantum network~\cite%
{QSTBook, QSTRevKimble, QSTRevBlatt}. Several techniques have been proposed
to implement robust QST in various physical systems, such as photon pulse
shaping of atoms coupled optical cavity~\cite{Zoller1997, Kuzmich2004},
transfer via spin chains and spin-wave engineering \cite{Bose2003,
Christandl2004, Sun2005, QSTRevBose, Yao2011}, frequency conversion via
optomechanical interface~\cite{TianAnnPhys}, and quantum error correction~%
\cite{Vermersch2017,Jiang2017}. However, the inevitable existence of
environmental noise and parameter imperfection can strongly limit the
fidelity of QST.

Topological phenomena, rooted in the global property of
topological matters, provide a natural protection against perturbation and
disorder~\cite{TopoRev1, TopoRev2}. Non-abelian anyons generated in
topological materials assisted with braiding operations have been
intensively explored for topological quantum computing~\cite{TQC, CooperTQC2}%
. The topologically protected Hall conductance is insensitive to disorder in
the electronic systems~\cite{IQH, TKNN}. Moreover, topologically protected
edge states can be used for robust disorder-immune photonic and phononic transport~\cite%
{Luling2014, Ozawa2018, Huber2016}. Recently, topological properties have
been employed for QST via two-dimensional chiral spin liquids and topological dipolar lattice \cite{Yao2013,Zoller2017}, which relies on the realization of
controllable coupling between qubits and the topological systems and is challenging to
implement. Therefore, it would be highly desirable to have a topologically
protected QST that can be implemented in practical qubit systems.

Here we present an experimentally feasible mechanism for implementing robust QST via the
topological edge states in superconducting qubit chains. By connecting
superconducting Xmon qubits into a one-dimensional chain with
tunable couplings~\cite{Martinis2013, Martinis2014,Roushan2017}, the generalized
Su-Schrieffer-Heeger (SSH) models~\cite{SSH, SSH2} are constructed, which
support various topological phases. We analytically derive the wave functions of the topological edge states in the above generalized SSH-type qubit chains and show that they have different forms of entangled states inside.  More importantly, via adiabatical ramping of the qubit couplings, we find that these topological edge states can be used as topologically-protected quantum channels to realize robust QST of single- and two-qubit entangled states. Using a numerical simulation, we quantitatively characterize the topologically protected robustness of the QST against qubit coupling disorders. Our result reveals that the QST is topologically protected by the finite energy gap between the bulk and the edge states and the transfer fidelities have a plateau at the value of unity in the presence of a sizable qubit-coupling imperfection. This protocol only requires tunable coupling between the qubits and can be implemented in various qubit systems, such as
trapped ions \cite{IonRev,Monroe}, cold atoms \cite{Bloch,Greiner}, nitrogen vacancy centers \cite{NVcentre_rev}, electronic spins \cite%
{Spin_rev} and optomechanical systems \cite{optomech_rev}.

This paper is organized as follows. In Sec. II, we present the generalized SSH-type qubit chains. In Sec. III and IV, we separately study how to transfer single-qubit states and two-qubit entangled states via the topological edge states in the $p=2$ and $p=3$ SSH-type qubit chains. We also investigate their robustness to qubit coupling imperfections. In Sec. V and VI, we give experimental discussions and a summary for our results.

\begin{figure}[htbp]
\centering
\includegraphics[width=8cm,height=5.2cm]{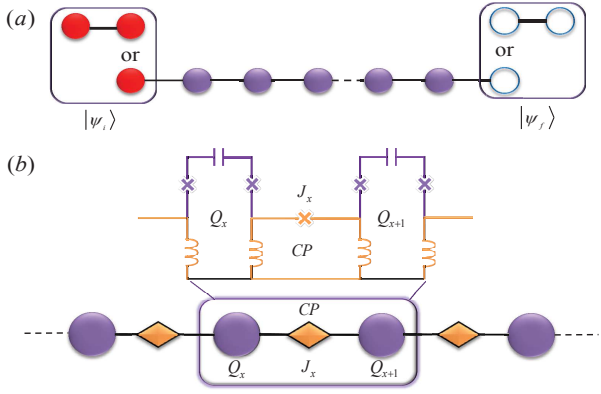}
\caption{(a) The transfer of unknown single-qubit or entangled states from
the qubits inside the left box to the qubits inside the right box through
the intermediate qubit chain. Each circle represents a qubit. (b)
The implementation of the qubit chain with superconducting Xmon qubits. The
qubits $Q_x$ and $Q_{x+1}$ are inductively coupled by the tunable coupler $CP$ with
coupling strength $J_x$. }
\label{fig1}
\end{figure}

\section{Generalized SSH-type qubit chains}

The generic setup for robust QST of single- or two-qubit states via the topological edge states in the generalized SSH-type qubit chains is illustrated in Fig.~\ref{fig1}(a). This protocol is applicable to various qubit systems, but for concreteness, here we focus on superconducting Xmon qubit chain \cite{QEC}. As shown in Fig.~\ref{fig1}(b), the coupling strength $J_{x}$ between adjacent Xmon qubits can be tuned smoothly by varying the current in the coupler \cite{Martinis2013, Martinis2014,Roushan2017}. The corresponding Hamiltonian for the Xmon qubit chain is
\begin{equation}
\hat{H}=\sum_{x}J_{x}\hat{\sigma}_{x}^{+}\hat{\sigma}_{x+1}^{-}+\text{H.c.},
\end{equation}
where $\hat{\sigma}_{x}^{+}=|e\rangle _{x}\langle g|$. We further let the tunable qubit coupling strength $J_{x}=g_{0}+g_{1}\cos(2\pi x/p+\theta)$, where $g_{0,1}$ are the coupling constants, $p$ is the number of qubits in one unit cell and $\theta $ is a control parameter. Such Xmon qubit chain generalizes the topological SSH model \cite{SSH,SSH2}. For $p=2$, each unit cell in the qubit chain has two qubits and the system can be described by a standard SSH model Hamiltonian with its topological phases characterized by winding numbers \cite{wnSSH}. For $p>2$, each unit cell has $p$ qubits and the qubit chain is described by a generalized SSH model Hamiltonian, where Chern numbers are employed to characterize the topological phases \cite{cnSSH}. According to the bulk-edge correspondence~\cite{TopoRev1, TopoRev2}, when the above qubit chains are in a topological phase, topological edge states exist in its boundaries. Since local perturbations cannot affect the properties of the bulk states, the topological invariants in these models will endow the edge states to be topologically protected against circuit imperfections. In this work, we will demonstrate that the edge states in the above $p=n$ SSH qubit chain can be employed as a topological quantum channel to robustly transfer $(n-1)$-qubit quantum state. Specifically, below we will consider the two cases $p=2$ and $p=3$. More importantly, we will also quantitatively investigate the effect of the coupling disorders on the transfer fidelity of single- and two-qubit states.

\begin{figure*}[htbp]
\includegraphics[width=13cm,height=4.5cm]{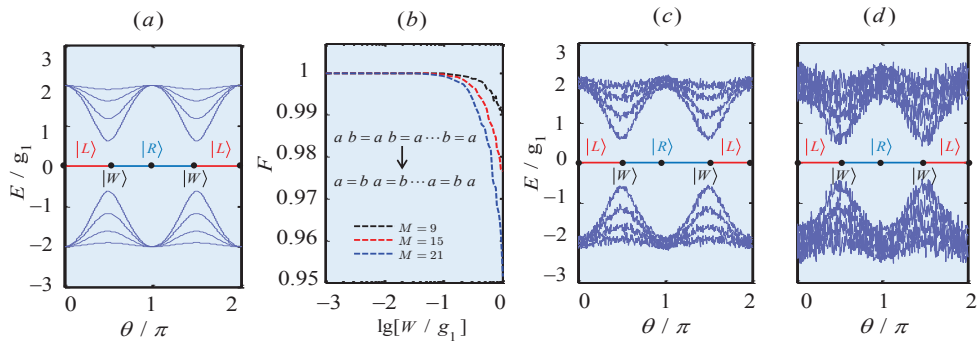}
\caption{The energy spectra of the $p=2$ SSH model vs $\protect%
\theta$ for the imperfection strength (a) $W=0$, (c) $W=0.6g_1$ and (d) $W=0.8g_1$. The total qubit number is $9$. (b) The fidelity of the QST vs the imperfection strength. The total qubit number is $9$ (the solid line), $15$ (the dashed line), and $21$ (the dash-dot line) with $\Omega=\{0.04g_1,0.02g_1,0.01g_1\}$. The other parameter is $g_0=g_1$.}
\label{fig2}
\end{figure*}

\section{Single-qubit quantum state transfer in $p=2$ SSH chain}

\subsection{Topological edge states in $p=2$ chain}

 Let us consider a $p=2$ SSH-type qubit chain with odd number ($2N-1$) qubits. Each unit cell contains two qubits labeled by $a$ and $b$, respectively. The resulted SSH-type qubit chain is described by the following Hamiltonian
\begin{equation}
\hat{H}=\sum_{x=1}^{N}(J_{1}\hat{\sigma}_{a_{x}}^{+}\hat{\sigma}%
_{b_{x}}^{-}+J_{2}\hat{\sigma}_{b_{x}}^{+}\hat{\sigma}_{a_{x+1}}^{-}+\text{%
H.c.}),  \label{SSH}
\end{equation}%
where $J_{i}=g_{0}+(-1)^{i}g_{1}\cos \theta $ ($i=1,2$) and $N$ is the
number of unit cells.  The edge states of a qubit chain with a single excitation are exponentially localized at the boundaries. The wave function of an edge state can be described by the following ansatz
\begin{equation}
|\psi _{E}(\theta )\rangle =\sum_{x=1}^{N}\lambda ^{x}(\theta )(\alpha
\sigma _{a_{x}}^{+}+\beta \sigma _{b_{x}}^{+})|G\rangle,
\label{ewSSH}
\end{equation}
where $|G\rangle=|gg\cdots g\rangle$ and the probability amplitude on site $x$ decays (increases) exponentially with the distance $x$ when $|\lambda |<1$ ($|\lambda |>1$), corresponding to the left (right) edge state, after the wave function is normalized. Let the eigenenergy of an edge state be $E$. Substituting (\ref{SSH}, \ref{ewSSH}) into the Schr\"{o}dinger equation $H|\psi _{E}\rangle =E|\psi _{E}\rangle $, we obtain
\begin{equation}
\begin{split}
E\lambda ^{x}(\alpha \sigma _{a_{x}}^{+}+\beta \sigma _{b_{x}}^{+})|G\rangle
&=[ J_{1}\lambda ^{x}(\beta \sigma _{a_{x}}^{+}+\alpha
\sigma _{b_{x}}^{+})\\
&+J_{2}(\beta \lambda ^{x-1}\sigma _{a_{x}}^{+}+\alpha
\lambda ^{x+1}\sigma _{b_{x}}^{+})] |G\rangle.
\label{eSSH}
\end{split}
\end{equation}%
It is straightforward to find that the edge state occupies only the $a$- or $%
b$-type qubits with $\beta =0$ or $\alpha =0$, respectively. The
corresponding eigenenergy is $E=0$.

For a qubit chain with odd number ($2N-1$) of qubits, the edge state energy spectrum is plotted in Fig. \ref{fig2}(a), with the topological edge state at zero energy and well separated from the bulk states. Specifically, we find that there is one edge state in the left end when $\theta \in (-\pi /2,\pi /2)$ and one edge state in the right end when $\theta \in (\pi /2,3\pi /2)$. Both the left and right edge states occupy the $a$-type qubits in each unit cell and are eigenstates of $\tau _{z}$, which leads to $\alpha =1$ and $\beta =0$ in (\ref{ewSSH}). This is because that the rightmost qubit is of $a$-type for a chain with odd number of qubits. With this analysis, we obtain $J_{1}+J_{2}\lambda =0$, i.e., $\lambda =-J_{1}/J_{2}$. The edge state wave function then can be derived as
\begin{equation}
|\psi _{E}\rangle =\sum_{x=1}^{N}(-1)^{x}\left[ \frac{%
g_{0}-g_{1}\cos (\theta )}{g_{0}+g_{1}\cos (\theta )}\right] ^{x}\sigma
_{a_{x}}^{+}|G\rangle,
\end{equation}%
which only occupies the $a$-type qubits. It can be easily verified that this edge state is localized near the left end when $\theta \in (-\pi /2,\pi /2)$ and $|\lambda |=|J_{1}/J_{2}|<1$, and is near the right end when $\theta \in (\pi /2,3\pi /2)$ and $|\lambda |=|J_{1}/J_{2}|>1$.

\subsection{Robust single-qubit quantum state transfer}

At $g_{0}=g_{1}$, the edge state concentrates towards the left (right) end when $\theta \in (-\pi /2,\pi /2)$ [$\theta \in (\pi/2,3\pi /2)$]. In particular, at $\theta =0$ and $\pi$, the coupling strength
becomes $J_{1}=0$ and $J_{2}=0$, respectively. The leftmost and rightmost qubits are decoupled
from the rest of the qubit chain. The edge states in this case become
\begin{eqnarray}
|L\rangle=|egg\cdots g\rangle\, \nonumber \\
|R\rangle =|gg\cdots ge\rangle.
\end{eqnarray}
At $\theta =\pi/2$ or $3\pi /2$, the edge state is a W-state $|W\rangle
=\sum_{x=1}^{N}(-1)^{x}\hat{\sigma}_{a_{x}}^{+}|G\rangle/\sqrt{N}$ with equal superposition of the excitations of all $a$-type qubits.

An unknown single-qubit state can be transferred adiabatically via the edge
mode. This is can be done by slowly ramping the qubit couplings to make $\theta$ varying linearly with time, i.e.,
\begin{equation}
\theta (t)=\Omega t,
\end{equation}
where $\Omega$ is the ramping frequency. Suppose $\theta $ is swept from $0$ at $t=0$
to $\pi $ at the final time. At time $t=0$, the leftmost qubit is prepared
in the unknown state $\alpha |e\rangle +\beta |g\rangle $ and all other
qubits are in their ground states. The state of the qubit chain is then $%
|\psi _{i}\rangle =\alpha |L\rangle +\beta |G\rangle $, which is in a
superposition of the edge state at $\theta =0$ and the ground state $%
|G\rangle $ with no excitation. When $\theta $ is varied from $0\rightarrow \pi /2\rightarrow \pi$,
the state evolves from $|L\rangle \rightarrow |W\rangle \rightarrow |R\rangle $,
then we realize the single-qubit quantum state transfer
\begin{equation}
|\psi _{i}\rangle =\alpha |L\rangle +\beta |G\rangle\longrightarrow|\psi_{f}\rangle =\alpha |R\rangle +\beta |G\rangle,
\end{equation}
where the rightmost qubit is in the state $\alpha |e\rangle +\beta |g\rangle $. To
ensure high fidelity of QST, it is required that the process be adiabatic in
the entire process, i.e., $\sqrt{g_{1}\Omega }$ needs to be smaller than the
energy gap between the bulk and the edge states. For example, we can choose $%
\Omega =0.01g_{1}$ for a chain of $21$ qubits, which has an energy gap
larger than $0.1g_{1}$. For superconducting Xmon qubits with $g_{1}/2\pi
=250$ MHz, the time of QST is $t_{f}=\pi /\Omega =0.2\,\mu \text{s}$, much
shorter than typical qubit decoherence times \cite{Martinis2013,Martinis2014}.

In practical, the system parameters cannot be perfectly tuned to
exact values due to the intrinsic fluctuations in device fabrication. In our
scheme, the main imperfection resides in the qubit coupling strengths, and it far exceeds the effect of qubit decoherence~\cite{Paik2011,Reagor2013}.
This imperfection can be described by the Hamiltonian
\begin{equation}
\hat{H}_{d}=\sum_{x}\delta J_{x}\hat{\sigma}_{x}^{+}\hat{\sigma}_{x+1}^{-}+%
\text{H.c.},
\end{equation}%
where $\delta J_{x}=W\delta $ with $W$ being the imperfection strength and $%
\delta \in \lbrack -0.5,0.5]$ being a random number. For each $\delta J_{x}$%
, we choose $100$ samples to perform the numerical simulation throughout
this work. The QST fidelity is obtained by averaging over the results of all
samples. Note that the time evolution of wave functions in the qubit chain is governed by the Schr\"{o}dinger equation $id\left\vert \psi (t)\right\rangle /dt=\hat{H}(t)\left\vert \psi
(t)\right\rangle $ with $\hat{H}(t)$ being the time-dependent Hamiltonian. The numerical simulation
of this evolution can be conducted via a fourth-order Runge-Kutta method.

In Fig.~\ref{fig2}(b), we numerically calculate the fidelity $F=|\langle R|\psi (t_{f})\rangle|$ as a function of the imperfection strength. A wide plateau at $F\approx 1$ appears for $W\lesssim 0.1g_{1}$, where the energy gap remains large enough to protect QST. The appearance of the plateau is a hallmark of the topologically assisted QST, which ensures high transfer fidelity, and the
plateau can also be observed in the two-qubit entanglement transfer studied
below. With current technology, the imperfection strength is $\sim 5\%$
of the coupling constant $g_{1}$. Our simulation shows that the fidelity can
exceed $0.998$ for $W=0.1g_{1}$ when the qubit chain size is over $20$. This
clearly demonstrates that nearly perfect QST can be achieved in practical circuits under
our protocol.

This topological protection is endowed by the chiral symmetry of this system. Such symmetry results in a symmetric energy spectrum with each positive eigenenergy $E$ accompanied by a negative eigenenergy $-E$, implying existence of zero energy edge mode. In the presence of qubit-coupling imperfection, the system Hamiltonian still obeys the chiral
symmetry, i.e., $\hat{\Gamma}(\hat{H}+\hat{H}_{d})\hat{\Gamma}^{-1}=-(\hat{H}+\hat{H}_{d})$,
where $\hat{\Gamma}=\prod_{x}(\hat{\sigma}_{a_{x}}^{+}\hat{\sigma}_{a_{x}}-%
\hat{\sigma}_{b_{x}}^{+}\hat{\sigma}_{b_{x}})$ is the chiral operator~\cite%
{SSH,SSH2}. As a result, the zero-energy edge state is insensitive to
imperfection in the couplings. This is verified by our numerical calculation in
Fig.~\ref{fig2}(c, d).

\begin{figure}[h]
\includegraphics[width=7cm, clip]{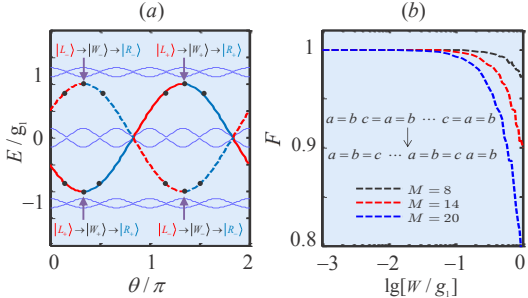}
\caption{(a) The energy spectra of the $p=3$ generalized SSH
model vs $\protect\theta $ with a chain of $8$ qubits and $g_{0}=0$.
(b) The fidelities of entanglement
transfer vs the imperfection strength. The total qubit number is $8$
(the solid line), $14$ (the dashed line), and $20$ (the dash-dot line) with $\Omega
=\{0.01g_{1},0.004g_{1},0.001g_{1}\}$, respectively.}
\label{fig3}
\end{figure}

\section{Entangled state transfer in $p=3$ SSH qubit chain}

\subsection{Topological edge states in $p=3$ chain}

In a $p=3$ generalized SSH-type qubit chain, each unit cell has three qubits labeled as $a$, $b$, and $c$, the corresponding Hamiltonian has the following form
\begin{equation}
\hat{H}=\sum_{x=1}^{N}(J_{1}\hat{\sigma}_{a_{x}}^{+}\hat{\sigma}%
_{b_{x}}^{-}+J_{2}\hat{\sigma}_{b_{x}}^{+}\hat{\sigma}_{c_{x}}^{-}+J_{3}\hat{%
\sigma}_{c_{x}}^{+}\hat{\sigma}_{a_{x+1}}^{-}+\text{H.c.}),  \label{gSSH}
\end{equation}
where $J_{s}=g_{0}+g_{1}\cos (2\pi s/3+\theta)$ ($s=1,2,3$) is the coupling strength.
As shown in Fig.~\ref{fig3}(a), there exists one pair of topological edge states in a $p=3$ SSH-type qubit chain with $3N-1$ qubits and $g_0=0$. Here one edge state exists within each bulk energy gap. The wave function of the edge states can be described by the following ansatz
\begin{equation}
|\psi _{E}(\theta )\rangle =\sum_{x=1}^{N}\lambda ^{x}(\theta )(\alpha
\sigma _{a_{x}}^{+}+\beta \sigma _{b_{x}}^{+}+\gamma \sigma
_{c_{x}}^{+})|G\rangle.  \label{ewgSSH}
\end{equation}
Denote the eigenenergy of an edge state as $E$. Substituting (\ref{gSSH}, \ref{ewgSSH}) into the Schr\"{o}dinger equation $H|\psi _{E}\rangle =E|\psi _{E}\rangle $, we obtain
\begin{equation}
\begin{split}
E\lambda ^{x}(\theta )(\alpha \sigma _{a_{x}}^{+}+\beta \sigma
_{b_{x}}^{+}+\gamma \sigma _{c_{x}}^{+})|G\rangle
=[J_{1}\lambda ^{x}(\beta \sigma _{a_{x}}^{+}+\alpha \sigma_{b_{x}}^{+})\\
+J_{2}\lambda ^{x}(\gamma \sigma _{b_{x}}^{+}+\beta \sigma
_{c_{x}}^{+})
+J_{3}(\lambda ^{x-1}\gamma \sigma _{a_{x}}^{+}
+\lambda
^{x+1}\alpha \sigma _{c_{x}}^{+})] |G\rangle.
\label{egSSH}
\end{split}
\end{equation}
Equation (\ref{egSSH}) can be solved for $\gamma =0$, where the $c$-type qubits are not occupied. Specifically, there are two eigenstates with $|\chi _{\pm }\rangle =(a_{x}^{\dag }\pm b_{x}^{\dag })/\sqrt{2}$ in unit cell $x$, and the coefficients $\alpha =1/\sqrt{2}$ and $\beta =\pm 1/\sqrt{2}$. Substituting these values into (\ref{egSSH}), we obtain the eigenenergies of the edge states $E_{\pm }=\pm J_{1}=\pm [ g_{0}+g_{1}\cos (2\pi /3+\theta )] $, which agree with the numerical result in Fig.~\ref{fig3}(a). Hence, there exist two branches of edge states, one in the upper and one in the lower energy gaps between the bulk states. Using (\ref{egSSH}), for the eigenstates $|\chi _{\pm}\rangle $, we derive $J_{2}\pm J_{3}\lambda =0$, i.e., $\lambda =\mp J_{2}/J_{3}$. The corresponding wave function of the edge states can be expressed as
\begin{equation}
|\psi _{\pm }\rangle =\sum_{x}\left[ \mp \frac{g_{1}\cos (4\pi
/3+\theta )}{g_{1}\cos \theta }\right] ^{x}\frac{\hat{\sigma}%
_{a_{x}}^{+}\pm \hat{\sigma}_{b_{x}}^{+}}{\sqrt{2}}|G\rangle,
\end{equation}
which only occupy the $a$- and $b$-type qubits.

\subsection{Robust two-qubit entangled state transfer}

The above edge states concentrate near the left end when $\theta \in (-\pi /6,\pi /3)\cup (5\pi /6,4\pi /3)$, and occupy the right end when $\theta \in (\pi /3,5\pi /6)\cup (4\pi /3,11\pi
/6) $. Specifically, at $\theta =\pi /6,\,7\pi /6$ and $\pi /2,\,3\pi /2$, the
coupling strength $J_{1}=0$ and $J_{2}=0$, respectively. In this case, the two leftmost and rightmost
qubits are decoupled from the rest of the qubit chain. The resulted edge states are
\begin{eqnarray}
|L_{\pm }\rangle =|\chi _{\pm }\rangle |gg\cdots g\rangle, \nonumber \\
|R_{\pm}\rangle =|gg\cdots g\rangle |\chi _{\pm }\rangle
\end{eqnarray}
where $|\chi _{\pm}\rangle =(|eg\rangle \pm |ge\rangle )/\sqrt{2}$ are Bell states.
At $\theta =\pi /3,\,4\pi/3$, the edge states are W states $|W_{\pm }\rangle =\sum_{x}(-1)^{x}(\hat{%
\sigma}_{a_{x}}^{+}+\hat{\sigma}_{b_{x}}^{+})|G\rangle/%
\sqrt{2N}$.

Suppose $\theta $ is swept linearly as $\theta(t)=\theta (0)+\Omega t$.
At time $t=0$, let $\theta (0)=\pi /6$, with the qubit chain prepared in the left edge
states $|L_{\pm }\rangle$, where the two leftmost qubits are prepared in the Bell state $|\chi _{+}\rangle$ and all other qubits are in their ground states. To prepare this state, we set the
frequencies of these two qubits to be far off resonance from the other
qubits, which effectively decouples these two qubits from the other qubits.
The Hamiltonian of this unit cell can thus be written as $\hat{H}_{0}=J_{0}%
\hat{\sigma}_{a}^{+}\hat{\sigma}_{b}^{-}+$H.c.. A driving pulse is then
applied to these qubits with the Hamiltonian $\hat{V}_{2}=\sqrt{2}\Omega _{0}\cos \left( \omega _{d}t\right) \left( \hat{\sigma}_{a}^{x}+\hat{\sigma}_{b}^{x}\right)$, where $\Omega _{0}$ and $\omega _{d}$ are the driving amplitude and frequency of the applied pulse, respectively. In the rotating frame of $\omega _{d}$, the driving pulse becomes $\hat{V}_{2}^{\text{rot}}=\Omega
_{0}(\hat{\sigma}_{a}^{x}+\hat{\sigma}_{b}^{x})/\sqrt{2}$. Let these qubits
be initially in the ground state $|gg\rangle $. With a driving frequency of $%
\omega _{d}=\omega _{q}+J$, the state $|\chi \rangle $ can be generated in a
duration of $t_{0}=\pi /2\Omega _{0} $. For $\Omega
_{0}/2\pi =100$ MHz, the operation time is $t_{0}=2.5\,\text{%
ns}$.

After initial state preparation, we adiabatically ramping the qubit couplings to sweep $\theta$. After a ramping time $t_{p}=\pi /3\Omega $, $\theta =\pi /2$. During the ramping, the state evolves adiabatically as $|L_{\pm }\rangle \rightarrow |W_{\pm }\rangle \rightarrow |R_{\pm }\rangle $, then we achieve the two-qubit entangled state transfer
\begin{eqnarray}
|L_{\pm }\rangle=|\chi _{\pm }\rangle |gg\cdots g\rangle \longrightarrow
|R_{\pm }\rangle=|gg\cdots g\rangle |\chi _{\pm }\rangle,
\end{eqnarray}
where the entangled state $|\chi _{\pm }\rangle $ is thus transferred
from the left end to the right end. With $g_{1}/2\pi =250$ MHz, we choose $%
\Omega =0.001g_{1}$ for a chain of $20$ qubits, which gives $t_{f}=0.67\,\mu
\text{s}$ and satisfies the adiabatic condition. We also numerically
simulate the transfer process in the presence of finite qubit-coupling
imperfection and obtain the transfer fidelity $F=|\langle R_{+}|\psi
(t_{f})\rangle |$ for the state $|\chi _{+}\rangle $. As shown in Fig.~\ref%
{fig3}(b), the fidelity exhibits a plateau at $F\approx 1$, demonstrating the
high robustness against qubit coupling imperfections. A fidelity above $0.99$
can be achieved for an imperfection strength $W\lesssim 0.07g_{1}$.

\begin{figure}[t]
\includegraphics[width=7cm,height=4cm]{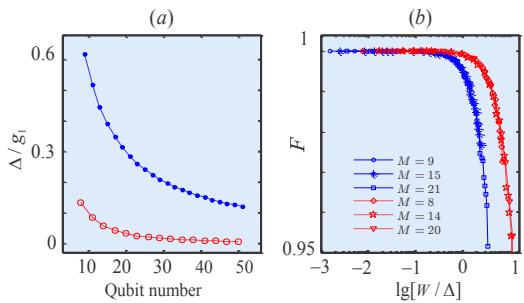}
\caption{ (a) The bulk-edge energy gap vs qubit number for the $p=2$ (point) and $p=3$ (circle)
SSH models. (b) The fidelities of the QST vs $\lg[W/\Delta]$. Blue:
the $p=2$ SSH model with $9$ (circle), $15$ (star), and $21 $ (square)
qubits; red: the $p=3$ generalized SSH model with $8$ (diamond), $14 $
(pentagram), and $20$ (triangle) qubits. Other parameters are the same as
those in Fig.~\protect\ref{fig2} and Fig.~\ref{fig3}.}
\label{fig4}
\end{figure}

\section{Discussions}

For the above QST protocols to succeed,
the adiabatic condition needs to be obeyed. Denote the energy gap
as $\Delta $, which is the smallest energy separation between the bulk and
the edge states in the related parameter range. The adiabatic theorem requires that $|dH/dt|< \Delta^{2}$. For the SSH models, this corresponds to $\sqrt{g_{1}\Omega}<\Delta$. The current state of art for superconducting circuits only can produce medium-sized superconducting quantum computer with qubit number $50-100$~\cite{Preskill2012, Boixo2016}. For a chain of $50$ qubits, $\Delta\sim g_1/10$ and a ramping rate $\Omega<0.01 g_{1}$ is required. When the qubit number is further increased, the gap near $\theta=0.5\pi$ will become much smaller, one can apply the shortcut-to-adiabaticity method \cite{STA,Du} to pass this point and realize the adiabatic quantum state transfer. One also can assemble many medium-sized qubit chains into a large-scale quantum network and use the topological edge states in each medium-sized qubit chain as quantum channels to realize a large-scale robust quantum state transfer.

Furthermore, we study the transfer fidelity of single-qubit state and entanglement as a
function of the parameter disorder $\lg [W/\Delta ]$. In Fig.~\ref{fig4}(b), the transfer fidelities for qubit chains with different size are plotted, which fall near a single curve for a given transfer regardless of the size of the chain size. Both curves have a wide plateau with
high fidelity exceeding $0.99$ when $W<0.1\Delta $. Our result verifies that
the QST via the edge states is topologically protected and insensitive to
small perturbations in the Hamiltonian.

Our system can be implemented with current technology of superconducting quantum devices. A chain of $9$ Xmon qubits~\cite{QEC} and a chain of $15$ flux qubits~\cite{flux} have been realized in experiments and the implementation of longer chains is promising in near future~\cite{Preskill2012, Boixo2016}. With a
typical coupling strength of $g_{1}/2\pi =250$ MHz, the ramping time for QST can be achieved in sub-micron seconds, much shorter than the decoherence times for the Xmon qubits \cite{Paik2011, Reagor2013}.

\section{Summary}

In summary, we have presented an experimentally realistic mechanism for implementing robust QST via topological edge states in superconducting qubit chains. The topological protected robustness of QST has been quantitatively demonstrated against qubit coupling imperfections with a numerical simulation. Our result indicates that high-fidelity QST between remote edge qubits can be achieved even in the presence of sizable qubit coupling imperfections. Our method can also lead to future studies of long-range edge-to-edge quantum entanglement~\cite{LDE} or scalable quantum networks with topologically protected quantum channels.

\section{Acknowledgment}

This work is supported by the National Key R\&D Program of China under Grants No.~2017YFA0304203 and No.~2016YFA0301803; the NSFC under Grants No.~11674200, No.~11604392, No.~11434007 and No.~91636218; the PCSIRT under Grant No.~IRT13076; SFSSSP; OYTPSP; SSCC; and 1331KSC. L.T. is supported by the National Science Foundation (USA)
under Award Numbers DMR-0956064 and PHY-1720501, the UC Multicampus-National
Lab Collaborative Research and Training under Award No. LFR-17-477237, and
the UC Merced Faculty Research Grants 2017.

\end{document}